\documentclass{ar-1col-S2O_rev}
\usepackage[numbers]{natbib}
\usepackage{url}
\jname{Author preprint of published article: Annu. Rev. Condens. Matter Phys.}
\jvol{17}
\firstpage{395}
\lastpage{418}
\jyear{2026}
\doi{10.1146/annurev-conmatphys-031424-010035}

\firstpagenote{Posted with permission from the Annual Review of Condensed Matter Physics, Volume 17$^\copyright$ by Annual Reviews, \url{http://www.annualreviews.org}.}

\usepackage{amsmath} 
\usepackage{amssymb}

\begin{document}

\markboth{Jensen \& Davis}{Physics of Soft Adhesion}

\title{The Physics of Soft Adhesion}

\author{Katharine E. Jensen$^1$ and Chelsea S. Davis$^2$
\affil{$^1$Department of Physics and Astronomy, Williams College, Williamstown, MA, USA, 01267; email: kej2@williams.edu}
\affil{$^2$Department of Mechanical Engineering and Department of Materials Science and Engineering, University of Delaware, Newark, DE, USA, 19716; email: chelsead@udel.edu}
}

\begin{abstract}  
This review provides an introduction to the essential physics of soft adhesion, 
including the thermodynamics of adhesion and wetting, the mechanics of contact with deformable materials, and the material properties that most affect interfacial interactions with soft solid gels and elastomers.
Throughout, we emphasize both foundational physics and current experimental and theoretical research in these areas.
We conclude with a practical overview of standard experimental test methods for characterizing soft adhesion. 
The physical understanding developed herein provides the basis for understanding the mechanics of contact with soft materials. 
\end{abstract}

\maketitle


\section{INTRODUCTION}
\begin{figure}[ht]
    \centering
    \includegraphics[width=\linewidth]{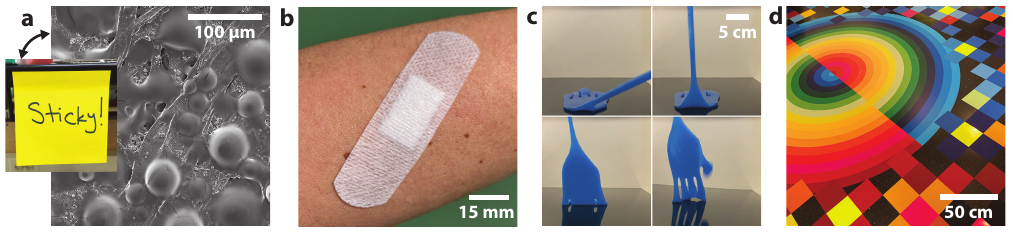}
    \caption{Everyday soft adhesion. 
    \textit{(a)} The adhesive on the back of a sticky note comprises thousands of $\sim 50$-$\mu$m-scale asperities that deform about a millimeter before debonding. 
    \textit{(b)} Soft adhesive medical bandages, tapes, and similar products revolutionized wound care by replacing stitches. 
    \textit{(c)} Soft ``sticky hands'' are popular toys. 
    \textit{(d)} Large-scale, removable adhesive wallpaper and floor coverings are increasingly used for decoration and advertising. 
    Electron micrograph in Panel \textit{a} provided by Nancy Piatczyc.
    Photographs in Panel \textit{c} provided by Meredith Taghon.
    All other photographs provided by Katharine E. Jensen.
    }
    \label{fig:soft_adhesive_examples}
\end{figure}

Soft adhesion abounds. 
In the course of just an hour, one might write on a sticky note, put a stamp on a letter, seal a box with tape, 
admire a large-scale, temporarily-adhered mural, 
or cover a wound with an adhesive bandage.
\textbf{Figure \textit{\ref{fig:soft_adhesive_examples}}} shows some examples of soft adhesion in action in the form of ``pressure-sensitive adhesives'' \cite{drew1930adhesivetape,creton2003pressure}. 
Even more ubiquitous than bandages and stickers, 
optically clear pressure sensitive adhesives bond the layers of a smartphone from the touchscreen down through deeper electronics \cite{abrahamson2020optically}, 
and temporary pavement marking tapes help guide cars safely through construction zones \cite{jo2021mechanical}.

All of these adhesives are made of soft matter, and a lot of soft matter is very sticky.
Although adhesive tape was first patented almost a hundred years ago \cite{drew1930adhesivetape} and a theoretical description of why more compliant materials are better able to establish adhesive contacts originated in the 1970s \cite{JKR1971},
research at the interface of soft matter physics and adhesion science has expanded significantly in recent years,
driven in part by discoveries of new physics via contact experiments with soft solids on small length scales \cite{Jerison2011,style2013universal,style2013surface}.

Soft matter exists at the intersection of traditional solids and fluids. 
This leads to unusual combinations of material properties, which in turn make soft materials desirable for a variety of engineering applications and a rich playground for discovering new physics.
Significant effort is being made to develop a quantitative understanding of how classic solid mechanics, particularly contact mechanics, changes for highly compliant solids;
how the unique structural and dynamic material properties of soft matter affect contact with these materials;
and how contact with a soft material may actually change its structure and properties.
The study and application of soft adhesion presents both opportunities and challenges that span a broad range of disciplines.

\begin{marginnote}[]
\entry{Pressure-sensitive adhesive (PSA)}{soft adhesive material that sticks after application of pressure, does not dry or cure (unlike a glue), and retains its surface and elastic properties so it can potentially be removed and reused}
\end{marginnote}

In this review, we introduce the essential physics of adhesive contact with soft, solid materials.  
Our aim is to provide an accessible guide for researchers new to the field of soft adhesion as well as for those who want to expand their understanding of fundamentals and recent developments across this broad, exciting field.
In Section \ref{sec:thermodynamics}, we describe the thermodynamics of surfaces that cause materials to be ``sticky,'' establishing quantitative connections between adhesion and wetting that allow us to consider contact with soft materials from these different but related perspectives. 
We next develop the fundamental mechanics of contact with compliant solids in Section \ref{sec:mechanics}, following a somewhat historical approach that simultaneously describes contact with softer and softer materials;
Section \ref{sec:how_soft} discusses how to determine what physics is expected to dominate at different stiffness and length scales. 
In Section \ref{sec:materials}, we focus specifically on the static and dynamic properties of soft solid gels and elastomers that affect and are affected by adhesion, as these are the 
broad classes of soft matter most commonly used in soft contact experiments and applications. 
Finally, in Section \ref{sec:test_methods},  we describe some experimental methods commonly used to characterize contact with soft materials.

\begin{textbox}[ht]\section{WHAT IS ``SOFT''?}
In this review, we follow the convention of the field of soft condensed matter physics and use the term ``soft'' as a synonym for ``compliant,'' and the opposite of ``stiff.''
We note that this differs from the other standard definition of ``soft'' as the opposite of ``hard,'' both referring to surface plasticity. 
While soft matter encompasses both compliant solids and fluids, here we focus on contact with solid materials, defined as having a measurable zero-frequency modulus. 
Wetting and adhesion with complex fluids also present exciting opportunities for research and discovery at the intersection of soft matter physics and contact mechanics, but are beyond our current scope.
\end{textbox}

\section{THERMODYNAMICS OF ADHESION: WHAT MAKES THINGS STICK?}
\label{sec:thermodynamics}

When two materials with different surface topographies meet, their mechanical properties determine whether they are able to form a conformal contact.
However, whether there is any energetic drive to create such an interface spontaneously depends on the thermodynamics of their surfaces. 
Two surfaces will adhere spontaneously only if doing so lowers the overall free energy of the system.
The following discussion assumes that external forces like gravity are negligible, and that surface interactions alone determine equilibrium. 

The pioneering work of Gibbs defined a surface (or interface) as a discontinuity in matter---that is, where the atoms or molecules of one material sharply give way to those of another---and sought to understand the associated surface and interfacial energies in terms of excess thermodynamic quantities \cite{gibbs1906scientific}. 
However, real material interfaces have some finite thickness;
rather than a true discontinuity between bulk and surface, atoms or molecules even a few layers into a material usually experience a different energetic environment than those deep in the bulk, which in turn contributes to the mean thermodynamic properties of the interface.
Gibbs thus noted that one can also define a surface as a two-dimensional (2D), geometric ``dividing surface'' (or 2D manifold) that is embedded in and everywhere parallel to the physical surface of the material \cite{gibbs1906scientific} (pg. 219).
This allows us to define thermodynamic properties of the dividing surface such that they are representative of the overall physical properties of the material surface.

\begin{marginnote}[]
\entry{Interface}{two-dimensional boundary between distinct materials where they come into contact}
\entry{Surface}{synonymous with interface, but usually refers to an interface between a condensed phase (solid, liquid) and either gas or vacuum}
\end{marginnote}

\subsection{Surface Energy $\gamma$ and Thermodynamic Work of Adhesion $W$}
\label{sec:gamma}

\begin{figure}[t]
    \centering
    \includegraphics[width=\linewidth]{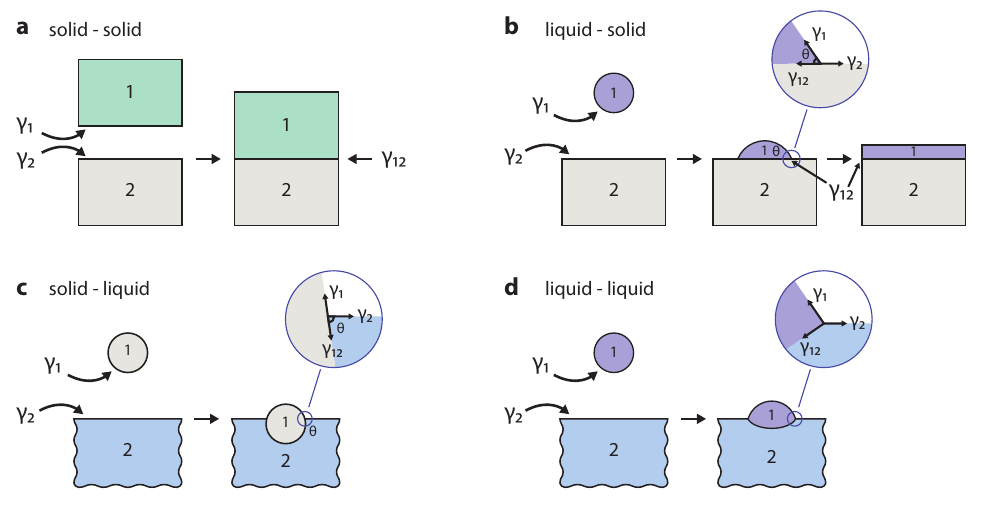}
    \caption{
    Adhesion and wetting both eliminate free surface areas with surface energies per area $\gamma_1$ and $\gamma_2$, respectively, in order to create an  interface with interfacial energy per area $\gamma_{12}$.
    (\textit{a}) Adhesion between two flat, solid materials. 
    (\textit{b}) Wetting between a liquid droplet (1) and a flat solid (2) results in partial or total wetting depending on the surface energy balance (Equation \ref{eqn:YD}) at the contact line \textit{(inset)}. 
    (\textit{c}) In contact between a rigid particle (1) and a fluid (2), partial wetting (\textit{inset}) results in adsorption, which adheres the particle to the liquid surface. 
    (\textit{d}) Wetting between a fluid droplet (1) and an initially-flat, immiscible fluid (2) results in deformation of both phases to create a Neumann triangle (\textit{inset}). 
    Illustration created by Roderick W. Jensen.
    }
    \label{fig:adhesion_wetting}
\end{figure}

The most important thermodynamic quantity for understanding contact mechanics is 
the surface energy, $\gamma$, which represents the free energy per unit area required to create the surface \cite{gibbs1906scientific} (pg. 315) \cite{deGennes2004,style2017elastocapillarity}.
For all condensed matter systems, $\gamma > 0$, and reflects the energetic penalty experienced by atoms or molecules on the surface of a solid or fluid compared to their counterparts in the bulk \cite{deGennes2004}.
When two flat, solid surfaces with areas $A$ and free surface energies $\gamma_1$ and $\gamma_2$, respectively, are brought into contact, the free surfaces are eliminated to create a new interface with the same area $A$ and an interfacial free energy per area $\gamma_{12}$, as shown in \textbf{Figure \ref{fig:adhesion_wetting}\textit{a}}.
In this simple example, the overall change in free energy to create this interface is 
\begin{equation}
    U_\text{adh}  =  \gamma_{12} A - (\gamma_1 + \gamma_2)A = -W A,
    \label{eqn:GibbsAdhesion}
\end{equation}
\noindent where $W$ is the thermodynamic work of adhesion per area, $W$, also called the adhesion energy, defined as:   
\begin{equation}
    W = \gamma_1 + \gamma_2 - \gamma_{12}
    \label{eqn:W_definition}
\end{equation}

When $W>0$, $U_\text{adh} < 0$, and creating the interface will lower the overall free energy.
In this case, adhesion is energetically favorable and will tend to happen spontaneously; 
the two materials will seem ``sticky'' to each other.
By contrast, if $W\leq0$, $U_\text{adh} \geq 0$, and an applied external force would be required to bring the surfaces into close contact.

\begin{marginnote}[]
\entry{example $\gamma$ values in mJ/m$^2$ (from \cite{deGennes2004})}{\\helium (4K): 0.1\\ cyclohexane: 25\\ glycerol: 63\\ water (100$^\circ$C): 58\\ water (20$^\circ$C): 73\\ mercury: 485\\ oil-water: $\gamma_{12}=50$}
\end{marginnote}

\subsection{Wetting as Adhesion}
\label{sec:wetting}

Solid adhesion and liquid wetting are closely related phenomena. 
This connection becomes even more important in contact with soft matter because highly compliant solids can exhibit significant liquid-like capillary mechanics  \cite{style2017elastocapillarity,tamim2024shaping}, including solid contact geometries that quantitatively resemble particle adsorption on fluid surfaces or liquid capillary bridges \cite{style2013surface,xujensen2017direct,jensen2017strain}.
Like adhesion, liquid wetting depends on the surface energy balance between maintaining a ``dry'' free surface and creating a ``wet'' interface \cite{deGennes2004}. 

To quantify wetting, we consider a small fluid droplet with surface energy $\gamma_1$ and negligible initial surface area spreading onto a flat, dry, rigid solid surface of area $A$ and initial dry surface energy $\gamma_2$, as shown in \textbf{Figure \textit{\ref{fig:adhesion_wetting}b}}.
The droplet can either spread to a finite extent, called partial wetting, or spread completely, eliminating the original dry surface and replacing it with both a wet interface with surface energy $\gamma_{12}$ and a fluid surface of equal area $A$.
This latter case, known as total wetting, results in an overall free energy change  
\begin{equation}
    U_\text{wet}  =  (\gamma_{12} + \gamma_1) A - \gamma_2 A = -S A,
    \label{eqn:GibbsWetting}
\end{equation}
\noindent where $S$ is the spreading parameter, related to the adhesion energy $W$, and defined as: 
\begin{equation}
    S = \gamma_2  - \gamma_{12} - \gamma_1 = W - 2\gamma_1
    \label{eqn:spreading}
\end{equation}
\noindent When $S\geq0$ (equivalently, $W \geq 2 \gamma_1$), the system is driven toward total wetting, while $S\leq -2\gamma_1$ (equivalently, $W \leq 0$) corresponds to non-wetting or mutually phobic surfaces.
In between, for $-2\gamma_1 \leq S \leq 0$ or $0 \leq W \leq 2\gamma_1$, 
the thermodynamic equilibrium droplet geometry in partial wetting (see \textbf{Figure \textit{\ref{fig:adhesion_wetting}b(inset)}}) is given by the Young-Dupr\'e Law  \cite{deGennes2004}, which can be expressed variously as: 
\begin{equation}
    \cos{\theta} = \frac{\gamma_2 - \gamma_{12}}{\gamma_1} = \frac{S}{\gamma_1} + 1 = \frac{W}{\gamma_1} - 1
    \label{eqn:YD}
\end{equation}
\noindent Here, $\theta$ is the equilibrium contact angle between a liquid droplet and a flat substrate in partial wetting \cite{deGennes2004,Gao2009Wetting}, as shown in \textbf{Figure \textit{\ref{fig:adhesion_wetting}b-c}}.
Total wetting corresponds to $\theta = 0$, while fully non-wetting or total phobia corresponds to $\theta = 180^\circ$.

Due to the quantitative connection between the thermodynamics of adhesion and wetting and the relevance of capillary mechanics to very soft solids \cite{style2017elastocapillarity,tamim2024shaping}, it is useful to introduce two other wetting configurations.
First, \textbf{Figure \textit{\ref{fig:adhesion_wetting}c}} inverts the wetting configuration of \textbf{Figure \textit{\ref{fig:adhesion_wetting}b}} to instead place a small, rigid, spherical particle of radius $R$ into contact with a flat fluid surface.
In this case, partial wetting results in particle adsorption, in which the particle is tightly bound to the interface at a depth $d$ set by satisfying the Young-Dupr\'e equation (see inset) and which scales as the particle radius, $d \sim R$.
Second, \textbf{Figure \textit{\ref{fig:adhesion_wetting}d}} demonstrates the geometry of contact between a fluid droplet and an initially-flat, immiscible fluid surface.
The equilibrium contact geometry is again set by a balance of surface tensions at the contact line (see inset), but 
with both materials free to change shape in response to surface forces, the equilibrium geometry takes the form of a Neumann triangle, balancing the surface tensions in all directions \cite{deGennes2004}. 

\begin{textbox}[h]\section{SURFACE ENERGY, SURFACE STRESS, OR SURFACE TENSION?}
In fluids, the surface free energy per area $\gamma$ is also referred to as the surface tension, as it results in tensile stresses that drive spontaneous motion to minimize surface area.
In solids, the picture is more complicated; 
unlike fluids, which are free to flow in response to stresses,  
it is possible to create new surface area in a solid either by bringing new molecules to the surface (e.g. cutting or plastic deformation), or by elastic stretching.
With this additional strain-dependent mechanism, the surface tension of a solid is more accurately represented by the surface stress tensor, $\mathbf{\Upsilon}$, which captures the total energetic cost per unit area of a free surface, including both the thermodynamic free energy per area $\gamma$ and energy stored in deformations of the surface \cite{gibbs1906scientific} (pg. 315) \cite{style2017elastocapillarity,tamim2024shaping,Shuttleworth1950}, and which may be strain-dependent \cite{Shuttleworth1950,cammarata1994surface,heyden2024distance}. 
Since soft materials can stretch significantly with relatively small forces, the energy contributions from surface stress can become increasingly important in ultrasoft contact. 
\end{textbox}

\section{SOFT CONTACT MECHANICS}
\label{sec:mechanics}

A positive adhesion energy $W>0$ tends to drive surfaces to stick together, but only if they can establish a non-zero contact area $A$.
If the individual free surfaces are not already perfectly matching, contact can occur only if one or both materials are sufficiently compliant to conform to the shape of the other.
Thus, in general, softer is stickier \cite{dahlquist1969criterion}, but it costs additional energy to deform an elastic solid.
This is the fundamental mechanics problem of adhesion with soft matter: 
adhesion energy drives deformation to create as much contact area as possible, yet restoring forces like elasticity and surface tension resist this deformation.
In this section, we articulate a theoretical mechanics description of adhesive contact that begins with relatively stiff materials and progresses to increasingly soft and complex surfaces, tracing the historic development of modern contact mechanics from the early work of Hertz 
to current theory encompassing adhesion, elasticity, and surface stresses.

\subsection{Foundations of Modern Contact Mechanics: from Hertz to JKR}

\begin{figure}[ht]
    \centering
    \includegraphics[width=\linewidth]{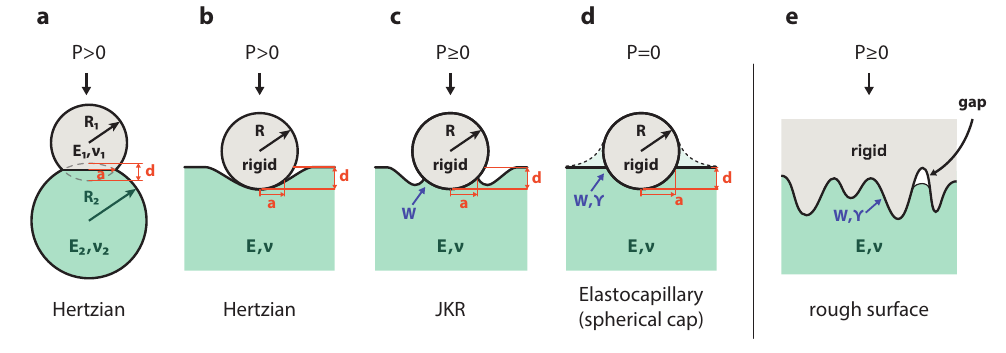}
    \caption{
    Adhesive contact geometries with increasing softness and interface complexity.
    Relevant parameters are indicated, including applied load $P$, sphere radius $R$, contact radius $a$, indentation depth $d$, Young modulus $E$, Poisson ratio $\nu$, adhesion energy $W$, and surface stress $\Upsilon$.
    \textit{(a-b)} Hertzian contact between \textit{(a)} two spheres 
    or \textit{(b)} a rigid sphere indenting an initially-flat, compliant elastic half space substrate due to an applied load $P>0$.
    \textit{(c)} A positive adhesion energy $W$ at the sphere-substrate interface drives deformation to create additional contact area and spontaneous indentation in JKR contact with or without an applied load.
    \textit{(d)} For highly compliant materials, solid surface tension $\Upsilon$ provides an additional restoring force that resists deformation of the substrate, resulting in adhesion that can quantitatively resemble capillary wetting (see also \textbf{Figure \textit{\ref{fig:adhesion_wetting}c}}). 
    Very soft substrates often have significant  out-of-plane deformation outside the contact area (dashed line, lightly shaded); 
    such ``adhesion ridges'' are often neglected in theoretical treatments of soft contact.
    \textit{(e)} Adhesion between a rough rigid surface and an initially-flat, compliant substrate that fails to make full conformal contact under the applied load, leaving a small gap. 
    Illustration created by Roderick W. Jensen. 
    }
    \label{fig:sphere-on-flat}
\end{figure}

In the late 1800s, Hertz established a continuum mechanics framework for describing the contact interactions between curved solid surfaces by combining continuum elastic theory with experimental data from pressing glass lenses together \cite{hertz1881beruhrung}, shown schematically in \textbf{Figure \textit{\ref{fig:sphere-on-flat}a}}.
This foundational work established how the contact radius $a$ between two somewhat deformable solid surfaces quantitatively depends on their respective isotropic elastic moduli (Young moduli: $E_1,E_2$, Poisson ratios: $\nu_1,\nu_2$), radii of curvature ($R_1,R_2$), and an applied loading force $P$ pressing them together.
Defining a reduced radius $R$ and an ``interface rigidity'' or ``effective modulus'' $K$ as 
\begin{equation}
    R = \frac{R_1 R_2}{R_1 + R_2} ~~ \text{and} ~~
    K = \frac{4}{3}\left(\frac{1-\nu_1^2}{E_1} + \frac{1-\nu_2^2}{E_2} \right)^{-1},
\end{equation}
the resulting generalized Hertz equations for $a$ and compression or indentation depth $d$ in the small deformation limit ($a \ll R$) are \cite{maugis2013contact}: 
\begin{equation}
    a_\text{Hz}^3 = \frac{R P}{K}; ~~ 
    d_\text{Hz} = \frac{a_\text{Hz}^2}{R} = \left(\frac{P^2}{K^2 R} \right)^{1/3} 
    \label{eqn:Hertz-a}
\end{equation}
We note that K is sometimes defined without the 4/3 prefactor (e.g. in Refs. \cite{lau2002spreading, carrillo2010adhesion}), and in this form is also commonly represented by the symbol $E^*$ (e.g. in Ref. \cite{shull2002contact}).

Depending on the application, the total Hertzian elastic energy $U_\text{Hz}$ from such a compression is commonly expressed in terms of load, geometry, or a combination of both \cite{maugis2013contact,JKR1971}: 
 
\begin{equation}
    U_\text{Hz} = \frac{2}{5} \frac{P^{5/3}}{R^{1/3} K^{2/3}}  = \frac{2}{5} P d = \frac{2}{5} \frac{P^2}{K a} = \frac{2}{5} \frac{K}{R^2} a^5 = \frac{2}{5} K R^{1/2} d^{5/2}
    \label{eqn:Hertz_energy}
\end{equation}

\noindent In the specific example geometry of a rigid sphere of radius $R$ indenting into a compliant half-space with uniform, isotropic elastic constants $E$ and $\nu$, shown in schematically in \textbf{Figure \textit{\ref{fig:sphere-on-flat}b}}, the reduced radius is identical to the sphere radius $R$, and $K$ is the rigidity of the compliant substrate alone, $K = \frac{4}{3} \frac{E}{1-\nu^2}$.

Hertzian mechanics neglects both frictional and attractive forces between the two contacting bodies, so deformation only occurs if $P>0$.
Johnson, Kendall, and Roberts (JKR) revisited the Hertzian contact problem in 1971 and added adhesive interactions at the interface, motivated by contemporary contact experiments with glass and rubber
that had measured ``considerably larger'' contact areas than predicted by Hertzian contact theory \cite{JKR1971}, as shown in \textbf{Figure \textit{\ref{fig:sphere-on-flat}c}}.
By explicitly incorporating the additional free energy change from creating a new interface with contact area $A = \pi a^2$, 
\begin{equation}
    U_\text{adh} = -W \pi a^2,
    \label{eqn:adhesion_energy_JKR}
\end{equation}
JKR analytically extended Hertzian contact theory to account for additional adhesion-driven deformation. 
For small-deformation sphere-sphere or sphere-on-flat contact with reduced radius $R$ and interface rigidity $K$, the general ``JKR Equation'' (adapted from Equation 19 in Ref. \cite{JKR1971}) is given by:
\begin{eqnarray}
    a_\text{JKR}^3 &=& \frac{R}{K} \left(P + 3 \pi W R + \sqrt{6 \pi W R P + \left(3 \pi W R \right)^2} \right) 
    \label{eqn:JKR-a} 
\end{eqnarray}
\noindent The adhesion-driven deformation also modifies the relationship between $a$ and $d$ in JKR contact to become \cite{shield1967load,maugis2013contact}:
\begin{equation}
    d_\text{JKR} = \frac{a_\text{JKR}^2}{3R} + \frac{2P}{3aK}
    \label{eqn:JKR-d}
\end{equation}

Two experimentally relevant results follow immediately from limiting cases of Equations \ref{eqn:JKR-a} and \ref{eqn:JKR-d} \cite{maugis2013contact,JKR1971,Maugis1995}.
First, in the absence of any applied external load ($P = 0$), JKR predicts a finite contact radius $a_0$ \cite{JKR1971} and spontaneous indentation $d_0$ \cite{maugis2013contact} due to adhesion:
\begin{equation}
    a_0^3 = \frac{6 \pi W R^2}{K} ~~ \text{and} ~~
    d_0 = \left( \frac{4 \pi^2 W^2 R}{3 K^2} \right)^{1/3}
    \label{eqn:JKR_zero_load}
\end{equation}
\noindent Equation \ref{eqn:JKR-a} additionally predicts that the tensile ``pull-off'' force required to separate the surfaces out of adhesive contact is independent of the elastic properties of the materials:
\begin{equation}
    P_\text{pulloff} = -\frac{3}{2} \pi W R
    \label{eqn:JKR_pulloff_force}
\end{equation}

Today, JKR theory and its adaptive extensions form the basis of modern contact mechanics.
They have facilitated a broad understanding of adhesion with soft surfaces. Thousands of studies have used JKR contact adhesion testing (see Section \ref{sec:JKR_test} below) to characterize the adhesive and elastic properties of diverse materials over a wide range of length scales \cite{shull2002contact,JohnsonBook1987}.
JKR theory and JKR tests \cite{shull2002contact} have been adapted to large deformations \cite{Maugis1995,maugis2013contact} and non-axisymmetric geometries \cite{chaudhury1996adhesive,barquins1988adherence} as well as for use with different materials and more complex interactions, including chemically reactive interfaces \cite{takahashi2008imaging}, patterned interfaces \cite{castellanos2011effect}, and interactions involving viscoelastic relaxation \cite{creton2003pressure,haiat2007approximate,carelli2007effect,josse2004measuring}. 

\subsection{Beyond JKR in Ultra-Soft Adhesion: Elastocapillary Contact Mechanics}
\label{sec:ecap}

As soft solids become even more compliant than the rubbers studied by JKR and contemporaries, other physical phenomena become increasingly important.
Notably, surface tension, traditionally only considered in fluid surface mechanics, can compete with or even dominate over bulk elasticity in very soft solids on small length scales.
A number of recent review articles have focused on various aspects of the rich and rapidly evolving field of ``elastocapillary'' mechanics  
\cite{style2017elastocapillarity,tamim2024shaping,bico2018elastocapillarity,andreotti2020statics}.
In the context of soft contact mechanics, liquid droplets on soft substrates cause significant deformation at the three-phase contact line, resulting in geometries that quantitatively resemble fluid-on-fluid rather than Young-Dupr\'e wetting \cite{Long1996, Jerison2011,style2013universal} (shown in \textbf{Figure \textit{\ref{fig:adhesion_wetting}d}} and \textbf{\textit{\ref{fig:adhesion_wetting}b}}, respectively).
Additionally, the predictions of JKR theory fail for highly compliant materials on small length scales due to capillary stresses, requiring further expansion of contact mechanics theory to include surface tension as an additional restoring force \cite{style2013surface,Salez2013,jensen2015wetting,cao2014adhesion,jensen2017strain,style2017elastocapillarity}.

In 2013, Style et al. \cite{style2013surface} expanded JKR theory in the specific case of sphere-on-flat, small deformation, and zero applied force ($P=0$) to include an additional free energy penalty from the stretching of the solid surface required to achieve conformal contact, finding the total energy to be: 
\begin{equation}
    U_\text{ecap} = \frac{2 \sqrt{3}}{5} K R^{1/2} d^{5/2} - 2 \pi W R d + \pi \Upsilon d^2
    \label{eqn:Style2013-energy}
\end{equation}
\noindent Here, $\Upsilon$ represents the solid surface tension or surface stress, taken to be isotropic and strain-independent but not necessarily the same as the surface or interfacial energy $\gamma$ \cite{style2013surface,xujensen2017direct,xu2018surface,schulman2018surface,heyden2024distance}.
The three terms of Equation \ref{eqn:Style2013-energy} represent contributions from Hertzian elastic contact, adhesion energy, and surface stress, respectively.

Taking the derivative of Equation \ref{eqn:Style2013-energy} with respect to indentation depth $d$, we obtain an implicit equation for $d$ as a function of $R$ in the absence of any applied force (adapted from Equation (2) in Reference \cite{style2013surface}):
\begin{equation}
    \frac{3}{\sqrt{3}} K R^\frac{1}{2} d^\frac{3}{2} - 2 \pi W R + 2 \pi \Upsilon d = 0
    \label{eqn:Style2013-force}
\end{equation}
This can be nondimensionalized \cite{jensen2015wetting} by defining a normalized radius $\bar{R}$ and a normalized indentation depth $\bar{d}$ given by 
\begin{equation}
    \bar{R} =  \frac{R}{\Upsilon/K} \left( \frac{W}{\Upsilon} \right)^{1/2}  ~~ \text{and} ~~
    \bar{d} = \frac{d}{\Upsilon/K} \left( \frac{W}{\Upsilon} \right)^{-1/2}, \\
\end{equation}
\noindent 
which simplifies Equation \ref{eqn:Style2013-force} to:
\begin{equation}
\frac{3}{2\pi \sqrt{3}} \bar{R}^{\frac{1}{2}} \bar{d}^{\frac{3}{2}} - \bar{R} + \bar{d} = 0
    \label{eqn:ecap_d_vs_R_nondim}
\end{equation}

This elastocapillary contact analysis uses a simplified geometry that approximates the contact geometry as a spherical cap \cite{weisstein2008spherical} indenting into an otherwise flat plane, such that $a^2 = 2 R d - d^2 \approx 2Rd$ for $d \ll a$.
The surface stress term is thus proportional to the difference in surface area between a disk of area $\pi a^2$ and a spherical cap with contact radius $a$ and indentation depth $d$, which is $\Delta A = \pi d^2$, and neglects the sometimes significant deformation outside of the contact line (shown schematically in \textbf{Figure \textit{\ref{fig:sphere-on-flat}d})}.
Nonetheless, these equations have been found to capture accurately the transition from JKR contact mechanics on larger length scales to capillary-dominated adhesion at small length scales, where adhesion quantitatively resembles particle adsorption on a fluid surface as shown in \textbf{Figure \textit{\ref{fig:adhesion_wetting}c}} \cite{style2013surface,jensen2015wetting,cao2014adhesion}.

Further studies have expanded this analysis via a combination of simulation and analytic approaches, including accounting for different experimentally relevant geometries, non-axisymmetric contacts, and applied external forces \cite{Salez2013,chakrabarti2018elastowetting,cao2014adhesion, hui2015indentation,ina2017adhesion,headley2025elastocapillary}.
However, accurate theoretical predictions of contact radius and/or indentation depth in elastocapillary adhesion experiments with an applied external load $P$ have proven very challenging \cite{hui2015indentation, liu2019effects}.
An elastocapillary equivalent of the full JKR Equation \ref{eqn:JKR-a} remains elusive.

\begin{textbox}[h]\section{DIMENSIONS AND UNITS OF SURFACE MATERIAL PROPERTIES}
All of the surface energy, stress, and tension properties described in this review (e.g. $\gamma$, $\Upsilon$, $W$, $S$, and $\mathcal{G}$) have dimensions of energy/area or, equivalently, force/length. 
Conventionally, energy/area units (e.g. mJ/m$^2$) are used for surface energies, while surface stresses and surface tensions are often given in force/length units (e.g. mN/m) to emphasize the connection with forces  on the interface. 
A consequence of these dimensions is that the ratio between any of these surface material properties and an elastic modulus or rigidity, e.g. $W/K$, is a length. 
Such length scales provide useful rules of thumb for estimating when surface properties begin to compete with elasticity in determining the overall mechanical response of soft materials.
For example, $\Upsilon/K$ provides an elastocapillary length scale for the transition from JKR to capillary-dominated adhesion and $\mathcal{G}_c/K$ is an elastoadhesive length that determines what scale of roughness can be spontaneously overcome.
\end{textbox}

\subsection{Determining Governing Physics via Competing Energy Scales}
\label{sec:how_soft}

As solid materials become more compliant, the governing physics for mechanical contact transitions from Hertzian elastic contact \cite{hertz1881beruhrung,boussinesq1885application} to JKR adhesive contact \cite{JKR1971,maugis2013contact} to the more wetting-like contact captured by elastocapillary theory \cite{style2013surface,Salez2013}.
What physics dominates at different stiffness and length scales can be determined by comparing each process's contribution to the overall contact energy.
In all cases, the contact theories described above are expected to break down at the molecular length scale ($R \sim 10^{-9}$), as the assumptions of continuum mechanics are no longer valid.

\begin{marginnote}[]
\entry{Energies per area or forces per length}{}
\entry{$\gamma$}{interfacial or surface energy (\S \ref{sec:gamma})}
\entry{$\Upsilon$}{interfacial or surface stress; $\Upsilon = \gamma$ for isotropic fluids}
\entry{$W$}{adhesion energy (Eq. \ref{eqn:W_definition})}
\entry{$S$}{spreading parameter (Eq. \ref{eqn:spreading})}
\entry{$\mathcal{G}$}{strain energy release rate (\S \ref{sec:fracture})}
\entry{$\mathcal{G}_c$}{critical strain energy release rate; $W \leq \mathcal{G}_c$ (\S \ref{sec:JKR_test})}
\end{marginnote}

\subsubsection{From Hertzian to JKR Contact Mechanics}

In order to understand when adhesion starts to cause significant deviations from classic Hertzian contact theory with an applied load $P$,  
we compare the contact energy $U_\text{Hz}$ given by Equation \ref{eqn:Hertz_energy} to the JKR adhesion energy $U_\text{adh}$ from Equation \ref{eqn:adhesion_energy_JKR} to obtain: 
\begin{equation}
    \frac{U_\text{Hz}}{U_\text{adh}} = \frac{2}{5 \pi} \frac{P^2}{K W a^3} = \frac{2}{5 \pi} \frac{P}{WR} 
    \propto \frac{P}{WR}
\end{equation}
The product $W R$ sets the force scale for crossover from load-dominated to adhesion-dominated JKR contact.  

In the case of contact with zero applied load ($P=0$), 
the competition between $U_\text{Hz}$ and $U_\text{adh}$ also determines when (at what values of $K$) spontaneous  
deformation of the interface will occur due to adhesion.
Comparing the Hertzian elastic deformation energy determined by contact geometry (Equation \ref{eqn:Hertz_energy}) to the adhesion energy we find:  
\begin{equation}
    \frac{U_\text{Hz}}{U_\text{adh}} = \frac{2}{5 \pi} \frac{K}{W} \frac{a^3}{R^2} = \frac{6 \sqrt{3}}{5 \pi} \frac{K}{W} \sqrt{\frac{d^3}{R}}
    \propto \frac{K}{W} \sqrt{\frac{d^3}{R}}
\end{equation}
We therefore expect the crossover rigidity at which spontaneous adhesion occurs to scale as $K \sim W \sqrt{R/d^3}$ \cite{abbott2015adhesion}.
For a rough surface with characteristic length scale $d \approx R$, 
the scaling simplifies to $K \sim W / R$.
Below this interface rigidity, a soft adhesive material would be expected to be able to make fully conformal contact without leaving open gaps as shown schematically in \textbf{Figure \textit{\ref{fig:sphere-on-flat}e}}.

A rule-of-thumb metric commonly used in both industrial and academic contexts for evaluating potential pressure-sensitive adhesives is the Dahlquist Criterion.
In order to be an effective adhesive material, the storage (shear) modulus (see Section \ref{sec:viscoelasticity}) measured at a frequency of 1 Hz should be $G' \leq 0.1$ MPa \cite{dahlquist1969criterion,dahlquist1966tack,gay2002stickiness}. Significant ``deterioration of tack'' \cite{dahlquist1966tack}(pg. 145) occurs above roughly $G' = 1$ MPa. 

One important limitation of JKR theory is that it is not accurate in the case of rigid solids on very small length scales, where instead the Derjaguin-Muller-Toporov (DMT) theory should be used  \cite{bradley1932lxxix,derjaguin1975effect,maugis1992adhesion,Maugis1995,greenwood2007dmt}.
The Tabor parameter $\mu = \left(16 R W^2/(9 K^2) \right)^{1/3}/\epsilon$, 
where $\epsilon$ describes a range of interaction of the surface forces, determines the transition between the JKR adhesion regime (at $\mu \gg 1$) and the DMT adhesion regime (at $\mu \ll 1$) \cite{tabor1977surface,maugis1992adhesion,greenwood1997adhesion,greenwood2007dmt}.
Adhesion with soft matter generally involves interactions with relatively large $W$ and small $K$, so DMT theory is rarely applicable.

\subsubsection{From JKR to Elastocapillary Adhesion}

Comparing the Hertzian elastic deformation energy to the capillary energy for indentation as a spherical cap to a depth $d$ we obtain:
\begin{equation}
    \frac{U_\text{Hz}}{U_\text{cap}} = \frac{2}{5 \pi} \frac{K}{\Upsilon} \sqrt{R d} \propto \frac{K}{\Upsilon} \sqrt{R d}
\end{equation}
We therefore expect elastocapillary mechanics to become dominant as the rigidity drops below about $K \sim \Upsilon/\sqrt{Rd} \sim \Upsilon/R$ if we again take the characteristic length scale of the deformations to scale as $d \sim R$, consistent with the spherical cap geometry \cite{weisstein2008spherical}.

\subsubsection{Toward Young-Dupré Wetting in the Low K Limit}
If the elastic modulus $K$ becomes so low that $U_\text{Hz}$ 
is negligible compared with $U_\text{adh}$ and $U_\text{cap}$, 
the limit of Equation \ref{eqn:Style2013-force} as $K \rightarrow 0$ predicts that $d/R = W/\Upsilon$, a dimensionless constant.
This is equivalent to spherical particles adsorbing to a fluid interface (see Section \ref{sec:wetting} and \textbf{Figure \textit{\ref{fig:adhesion_wetting}c}}), in this case with a constant macroscopic contact angle between the particle and the far-field flat surface, $\theta = \cos^{-1}(W/\Upsilon - 1)$.
Multiple experiments with spherical particles on highly compliant substrates have directly observed such adsorption-like adhesion \cite{style2013surface,cao2014adhesion,jensen2015wetting}, and the approach to this limit was recently demonstrated experimentally with the inverted geometry of highly compliant microspheres adhered to flat, rigid substrates \cite{headley2025elastocapillary}.

\subsection{Adhesion as Fracture} 
\label{sec:fracture}

We have developed our discussion of soft adhesive mechanics primarily from a thermodynamic perspective, using an energy balance to find equilibrium contact geometries.
An alternative approach, developed by Maugis et al. \cite{maugis2013contact,maugis1978fracture,barquins1982adhesive,maugis1992adhesion}, instead treats adhesion as a fracture mechanics problem 
by building on continuum mechanics literature that analytically solved the ``Boussinesq'' problem of a rigid, axisymmetric indenter of arbitrary profile indenting an elastic half space \cite{boussinesq1885application,sneddon1965relation,maugis1992adhesion}.

In this framework, creating an interface between two materials is equivalent to closing a crack; 
adhesion can thus be directly expressed in terms of fracture mechanics quantities like the strain energy release rate $\mathcal{G}$, equal to the energy cost per unit interface area required to open the crack \cite{griffith1921vi,maugis1992adhesion}.
In equilibrium, $\mathcal{G} = W$;
more generally, $\mathcal{G} \leq W$ for an advancing contact (increasing indentation depth $d$) and $\mathcal{G} \geq W$ for a receding contact, because measurements of $\mathcal{G}$ represent a rate-dependent, mechanical rather than thermodynamic balance that includes the work done in deforming the contacting materials \cite{shull2002contact}.
Adhesion measurements based on JKR adhesion theory (see Section \ref{sec:JKR_test}) generally measure $\mathcal{G}$ and identify a critical strain energy release rate $\mathcal{G}_c$ as an upper bound for $W$.

One advantage of framing adhesion as fracture is that the analysis is more readily expanded to a broader array of contact geometries than the thermodynamic approach, including larger deformations \cite{Maugis1995} and non-axisymmetric geometries like cylindrical contact \cite{barquins1988adherence,barquins1998etat,chaudhury1996adhesive,baney1997cohesive}.
An additional advantage is that one can predict the equilibrium contact geometry both inside and outside the contact radius and also investigate the stability of the contact \cite{maugis2013contact,maugis1978fracture}.
We refer readers interested in learning more about adhesion and fracture of soft materials to the comprehensive References \cite{maugis2013contact,creton2016fracture}.

\subsection{Soft Adhesion to Rough Surfaces}
\label{sec:rough}

Real solid surfaces exhibit roughness over many length scales, from bumps easily visible to the naked eye to microscopic asperities down to the atomic scale.
While surface roughness can be quantified in many ways, a standard method in both adhesion and tribology is to calculate the power spectral density of the surface heights, $C(q) =  |\tilde{h}(\textbf{q})|^2$, where $\tilde{h}(\textbf{q})$ is the Fourier transform of the surface height and $\textbf{q}$ is a wave vector \cite{muser2017meeting,muser2022modeling}.
Accurately measuring and understanding the roughness of surfaces even as seemingly smooth as nanocrystalline diamond is very challenging \cite{jacobs2017quantitative,gujrati2018combining,dalvi2019linking,pradhan2025surface};
recent work has even proposed to treat surface topography as a new material parameter \cite{jacobs2022surface}.

Surface roughness can dramatically affect the strength of an adhesive interface, depending on whether the adhesive is soft enough to deform into conformal contact without leaving significant gaps in contact, as shown schematically in \textbf{Figure \textit{\ref{fig:sphere-on-flat}e}}.
Surface roughness can enhance \cite{greenwood1981mechanics} or limit adhesion with soft materials, depending on surface energies, material stiffness, and the scale of roughness.
Using a fracture mechanics approach to adhesive contact, Persson and Tosatti developed a framework to account for both the energetic cost $U_\text{el}$ of elastic deformation to conform to a rough surface and the additional contact area gained by making conformal contact across a rough topography, yielding an apparent adhesion energy $W_\text{app}$ that may be larger or smaller than the thermodynamic work of adhesion $W$ \cite{persson2001effect,persson2022functional}. 
One of the enduring challenges in understanding soft adhesion with rough surfaces is accurately measuring or calculating the real area of contact, which has recently inspired creative efforts to address this systematically \cite{muser2017meeting,style2018contact,pradhan2025surface}.
Understanding how surface roughness affects the dynamics of adhesive contact formation and interfacial separation is an area of active research \cite{sanner2024soft}.

In applications, researchers have taken inspiration from the biological features observed on the toe pads of geckos, spiders, ants, and beetles \cite{arzt2003micro} and leveraged surface roughness to vary adhesive strength \cite{Sitti2003SynGecko}, develop materials with tunable adhesive properties \cite{deneke2021pressure}, or modify adhesion using wrinkled or periodically buckled surface structures \cite{Davis2012Enhanced}.

\section{HOW SOFT MATERIAL PROPERTIES INTERACT WITH ADHESION}
\label{sec:materials}

Soft adhesion requires soft materials. 
Soft solids generally will be stickier than their stiffer counterparts because they can conform to establish larger contact areas, but real soft matter generally comprises quite complex materials whose unique material properties both enrich and complicate our picture of soft adhesion. 
For the purposes of this review, we focus on the most common classes of soft matter used in contact mechanics research and applications: networked polymer gels and elastomers.
Although the full scope of pressure sensitive adhesives is larger than this, and soft matter more generally is truly vast, studying gels and elastomers both enables insight into new physics discovered in such soft materials and provides a framework for understanding adhesion with broader classes of soft matter.

\subsection{Structure and Properties of Soft Gels and Elastomers}

Networked solid polymer materials usually consist of long-chain polymers that are crosslinked to form a percolated, system-spanning network.
Such solids have very high material compliance compared with traditional engineering materials like metal and glass because polymers largely accommodate elastic deformations entropically, via extension and straightening of long polymer chains, rather than enthalpically via stretching of interatomic bonds, assuming that the polymer is well above its glass transition temperature  \cite{doi1988theory,binder2011glassy,rubinstein2003polymer,andreotti2020statics}.
The polymers themselves can have a variety of architecture, from linear chains to highly branched structures.
The connectivity of the network, created by crosslinking the polymers together, generates the overall solid elasticity of the solid, with higher crosslinking densities corresponding to stiffer material \cite{doi2013soft}.

\begin{figure}
    \centering
    \includegraphics[width=\linewidth]{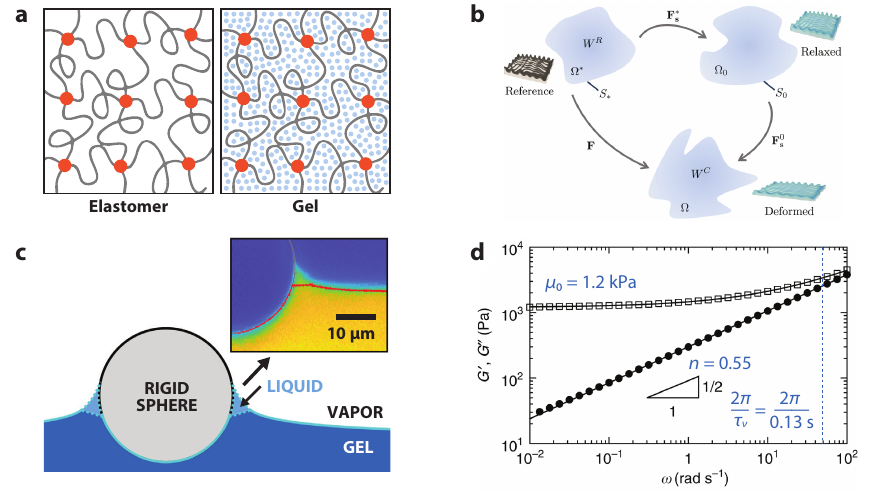}
    \caption{
    Static and dynamic properties of soft matter affect adhesive contact.
    \textit{(a)} Schematic comparison of elastomer versus gel structure.
    Both have system-spanning, crosslinked elastic networks (gray lines linked by red circles);
    in gels, the network is also permeated by a continuous free fluid phase (blue dots). 
    \textit{(b)} Surface and bulk properties of soft materials can be highly strain-dependent, affected by both fabrication pre-strains and deformations during experiments \cite{heyden2024distance}. 
    \textit{(c)} Adhesion and wetting can induce local fluid phase separation in soft gels, modifying both the equilibrium structure and dynamics of making an breaking contact \cite{jensen2015wetting, cai2024phase};
    confocal microscopy image \textit{(inset)} shows the fluorescently-dyed gel fluid phase (yellow) establishing a sharp wetting ridge beyond the surface of the gel elastic network (red dots).
    \textit{(d)} Soft rubbery gel materials like silicones typically demonstrate power law rheology. 
    Lines show fit to Equation \ref{eqn:chasset-thirion}, with fit parameters noted.
    Panel \textit{a} illustration created by Roderick W. Jensen. 
    Panel \textit{b} reproduced from Reference \cite{heyden2024distance} with permission from the Royal Society of Chemistry. 
    Panel \textit{c} adapted from Reference \cite{jensen2015wetting}; 
    Panel \textit{c (inset)} data provided by Katharine E. Jensen. 
    Panel \textit{d} adapted from Reference \cite{karpitschka2015droplets}. 
    }
    \label{fig:soft_matter_adhesion}
\end{figure}

Elastomers and gels represent two major categories of network polymer systems, shown schematically in \textbf{Figure \textit{\ref{fig:soft_matter_adhesion}a}}.
Elastomers, like rubber, consist of interconnected networks of large molecules that are fully crosslinked, either physically or chemically, without additional dangling or free chains. 
Polymeric gels, by contrast, are heterogeneous materials consisting of a percolated elastic network permeated by a continuous fluid phase that is free to diffuse and flow through the elastic network.
The fluid phase can be chemically the same as the elastic network, but uncrosslinked, or a different material altogether, as in the case of hydrogels and biopolymer gels where the fluid phase is primarily water.
Recent reviews have examined the structure and properties of gels and elastomers in relation to soft wetting \cite{andreotti2020statics}, their mechanical properties \cite{mckenna2018soft}, and competing gel relaxation mechanisms \cite{hu2012viscoelasticity}.
Notably, the terminology describing soft networked polymers is not yet fully standardized; even some fairly recent papers described polymer gels generically as elastomers because the importance of the structural distinction was not yet appreciated \cite{style2013surface,hourlier2017role}.

Most of the very softest solid polymer materials are gels.
While fully crosslinked rubber elastomers can easily be formulated to have Young's moduli $E \gtrsim 1$ MPa \cite{gent2006mechanical}, the presence of uncrosslinked chains or another fluid phase is usually required to achieve moduli $E \lesssim 100$ kPa.
For example, many commonly used PDMS gels and elastomers are long-chain linear polymers partially crosslinked via reactive end groups \cite{lisensky1999replication}, in which both the fluid fraction and elastic stiffness can be modified by changing the crosslink density or by swelling a polymer gel with different fluids \cite{glover2020extracting}.
Recently developed bottlebrush polymer architectures offer new opportunities for decoupling gel fluid content and elastic stiffness modulus, including some solvent-free elastomer formulations with sub-kPa moduli \cite{cai2015soft,cao2015computer,daniel2016solvent}. 
Gels engineered to have a crosslinking gradient have been shown to be superlubricious in friction experiments \cite{chau2023designing}.
Overall, there exists a rich material design space of networked polymer architectures and compositions that remains open for systematic exploration, especially in terms of understanding the coupling of material properties to surface mechanics. 

\subsubsection{Softness through geometry}
\label{sec:geometry}

This review focuses on materials that are inherently soft by virtue of their bulk elastic moduli, that is, having very low material compliance.
However, we note that it is also possible to achieve high mechanical compliance via micro- or meso-scale structuring of a stiff material with a very high Young modulus;
this ``geometric softness'' or ``geometric compliance'' is commonly used by biological organisms like geckos to achieve strong, reversible adhesion \cite{Autumn2002Gecko, arzt2003micro, Sitti2003SynGecko}.
While a thorough discussion is beyond the scope of this review, the study and development of adhesive contact with geometrically soft materials---especially those that are bio-inspired---is a rich and highly active field that has been addressed in other recent review articles \cite{brodoceanu2016hierarchical,favi2014inspiration}.

\subsection{Strain-Dependent Bulk and Surface Material Properties}
\label{sec:strain_dependent}

Many material properties, including both bulk elasticity and surface stress, are expected to be strain-dependent outside of very small deformations \cite{fu2001nonlinear,cammarata1994surface}. 
Unlike traditional engineering materials like metals and glass that yield at only a few percent strain \cite{telford2004case}, soft solids like gels and elastomers can often be deformed elastically to as much as hundreds of percent strain without yield or fracture \cite{lake1967strength,ahagon1975threshold, gent2001rubber}.
This complicates our fundamental physical understanding of these materials by requiring that we account for strain-dependent material properties in quantitative descriptions of their contact mechanics, but also opens up new opportunities to use soft matter to investigate strain-dependent material properties and mechanical responses that were previously inaccessible in much stiffer materials.

\subsubsection{Strain-Dependent Elasticity}

Nearly all materials exhibit non-linear elasticity outside of small deformations \cite{love2013treatise,treloar1975physics}, 
including rubber and other elastomers and gels \cite{mooney1940theory,rivlin1948large,gent2001rubber,GENT20051,dobrynin2011universality,carrillo2013nonlinear}.
Numerical models of network polymer gels and elastomers at large deformations often employ entropic nonlinear elastic models \cite{dobrynin2011universality,carrillo2013nonlinear} and/or treat the materials as neo-Hookean incompressible solids (see e.g. \cite{liu2019effects}). 
At the same time, many soft solids exhibit linear elasticity to much larger strains, even to tens or hundreds of percent deformation, 
and in the case of extremely low elastic stiffness, the contribution to the overall energy from elastic deformations can be so small compared to surface effects that the precise form of the elastic constitutive relationship becomes insignificant \cite{headley2025elastocapillary}.
Thus, approximating elasticity as strain-independent may be appropriate for soft matter, even when other material properties exhibit non-negligible strain dependence \cite{xujensen2017direct,jensen2017strain}.

\subsubsection{Strain-Dependent Surface Properties}

While the surface energy $\gamma$ is a thermodynamic scalar quantity, the surface stress (or surface tension) $\Upsilon_{ij}$ of a solid is a two-dimensional surface tensor equal to the work required to create a new unit of surface area by stretching the surface by an applied strain $\epsilon_{ij}$.
This difference was originally pointed out by Gibbs (see Ref. \cite{gibbs1906scientific}, pg. 315) and has been discussed in detail in more recent reviews \cite{cammarata1994surface,style2017elastocapillarity}. 
The two quantities can be related to each other by considering a reversible deformation $d \epsilon_{ij}$ applied to a surface, yielding the Shuttleworth equation 
\cite{Shuttleworth1950,gurtin1978surface,cahn1980surface,cammarata1994surface,style2017elastocapillarity,heyden2024distance}: 
\begin{equation}
    \Upsilon_{ij} = \gamma \delta_{ij} + \frac{\partial \gamma}{\partial \epsilon_{ij}}
    \label{eqn:shuttleworth}
\end{equation}	
Here, $\delta_{ij}$ is the Kroniker delta, and both $\gamma$ and $\Upsilon_{ij}$ refer to the current state of the system rather than a fixed reference state \cite{cahn1980surface,style2017elastocapillarity}.
To leading order in surface shear and dilational strain components, the surface stress for an isotropic material can be expressed in terms of surface elastic constants as \cite{style2017elastocapillarity,xu2018surface}
\begin{equation}
    \Upsilon_{ij} = \Upsilon_0 \delta_{ij} + \lambda^s \epsilon^s_{kk} \delta_{ij} + 2 \mu^s \epsilon^s_{ij}
\end{equation}
\noindent where $\Upsilon_0$ is the zero-strain scalar surface tension, $(\lambda^s,\mu^s)$ are surface Lam\'e constants, and the repeated index indicates summation.

In liquids, it is not possible for the surface to remain indefinitely in a state of strain, and thus the surface tension is identical to the surface energy, $\Upsilon = \gamma$. 
In solids, however, the surface tension is not generally the same as the surface energy \cite{gibbs1906scientific,cammarata1994surface}. 
This has long been understood to have important consequences for the mechanics of materials like metals \cite{cammarata1994surface}, but whether and to what degree the Shuttleworth equation applies to soft matter systems remains a subject of vigorous study and debate \cite{xujensen2017direct,liu2019effects,xu2018surface,liang2018surface,liang2018surface2,jensen2017strain,schulman2018surface,andreotti2016soft,masurel2019elastocapillary,bain2022reply,heyden2024distance}.
Very recent theoretical work has sought to reconcile inconsistencies in the existing literature and offer a new mathematical framework, derived from finite kinematics and continuum elasticity, that explicitly accounts for large deformations and, critically, establishes the appropriate choice of reference state from which to measure stress and deformation, as shown in \textbf{Figure \textit{\ref{fig:soft_matter_adhesion}b}} \cite{heyden2024distance}.

In the context of contact mechanics, recent experiments and simulations have used adhesive contact to measure strain-dependent surface stress \cite{xujensen2017direct}, and investigated the connection between strain-dependent surface stresses and the stiffness of adhesive contacts with PDMS gels \cite{liu2019effects,jensen2017strain,wu2018effect}, but key discrepancies persist between experimentally observed contact geometries and simulation results. 
The possible strain dependence of other surface properties such as adhesion \cite{fretigny2017contact,zheng2017indentation} is far less investigated and is an area of active study.

\subsection{Adhesion-Induced Phase Separation}

Generally, a gel's fluid phase remains within the elastic network, and gels behave macroscopically very similarly to homogeneous elastomers well above the molecular scale.
However, local deformations from contact can cause the internal fluid phase to separate from the gel elastic network, with significant consequences for wetting and adhesion \cite{jensen2015wetting}.
For example, aqueous droplets sliding on a PDMS substrate become lubricated by extracted oil \cite{hourlier2017role}, an example of adaptive wetting \cite{wong2020adaptive,hauer2024wetting}.

In contact mechanics, describing the precise nature of the contact line has long presented a mechanical conundrum because forming a sharp cusp in an elastic solid as predicted by JKR and related adhesion theories would require an elastic stress singularity.
Gels are able to avoid this elastic stress singularity by undergoing a local adhesion- or wetting-induced phase separation at the contact line, as shown in \textbf{Figure \textit{\ref{fig:soft_matter_adhesion}c}}, releasing some of their fluid phase to mediate deformation near contact.
Instead of the three-phase contact line predicted by classic contact mechanics theory (\textbf{Figure \textit{\ref{fig:sphere-on-flat}a-d}}), the gel instead forms a four-phase contact zone with two additional, hidden contact lines at the boundaries between the gel and its extracted fluid phase \cite{jensen2015wetting}.

Contact-induced phase separation has been shown to affect the forces and geometry of gel adhesive contact \cite{jensen2015wetting,pham2017elasticity,glover2020capillary,headley2025elastocapillary} and wetting \cite{qian2024emergence}, as well as the sliding dynamics of droplets contacting a soft gel surface \cite{hauer2023phase,jeon2023moving} and even more general ``osmocapillary'' phase separation phenomena \cite{liu2016osmocapillary,liu2025osmocapillary}.
Understanding the fundamental coupling between mechanics and phase in soft solid materials remains an area of very active research, especially given the dynamic nature of this interplay; 
the structure and composition of the material determine its contact mechanics, but simultaneously the contact can modify the material properties.
Whether there is a non-zero surface tension $\Upsilon_{gl}$ between a gel and its own fluid phase also remains an open question \cite{jensen2015wetting,pham2017elasticity,glover2020capillary}, as does that of what role this phase separation may play in mediating contact with more complex rough surfaces, in terms of both contact area and adhesive forces.

\subsection{Competing Relaxation Mechanisms: Viscoelasticity and Poroelasticity}

Polymer architecture and network composition are important determiners of the inherent relaxation time scales in soft gels and elastomers \cite{hu2012viscoelasticity,doi2009gel,andreotti2020statics}, which in turn determine dynamic behavior in adhesive contact.
Both elastomers and gels can undergo stress relaxation and energy dissipation via viscoelastic processes, which involve rearrangement of the gel elastic network structure over time.
Because of their internal fluid phase, gels have an additional, competing relaxation mechanism via poroelastic flow, in which the internal fluid phase of the gel flows relative to the gel elastic network structure.
Both viscoelastic and poroelastic relaxation are important in governing gel deformation dynamics, and thus need to be considered in understanding dynamic adhesion with soft matter.

\subsubsection{Gel and Elastomer Viscoelasticity: Rubbery Rheology}
\label{sec:viscoelasticity}

Many polymer gels and elastomers exhibit power law rheology, in which at high applied frequencies $\omega$, both the storage (elastic) modulus $G'(\omega)$ and the loss (viscous) modulus $G''(\omega)$ scale as $G' \sim G'' \sim \omega^n$.
The power law exponent $n$ is a constant that that depends on the polymer architecture, usually between 0.5 and 2/3 for gels and elastomers \cite{andreotti2020statics,karpitschka2015droplets,chambon1985stopping,scanlan1991composition,martin1988viscoelasticity,chasset1966viscoelastic,curro1983theoretical,Long1996}.  
The rheological response of such materials are well fitted by the Chasset-Thirion model, which was originally developed to describe the viscoelastic response of rubbers over time during creep experiments \cite{chasset1966viscoelastic}. 
In its modern form \cite{karpitschka2015droplets}, the Chasset-Thirion equation is expressed in terms of the complex shear modulus $G^*$ as  
\begin{equation}
    G^*(\omega) = G'(\omega) + iG''(\omega) = \mu_0 \left(1+ (i \omega \tau_v)^n \right)
    \label{eqn:chasset-thirion}
\end{equation}
\noindent where $\mu_0$ is the zero-frequency shear modulus and $\tau_v$ is the viscoelastic relaxation time.
Typical frequency-sweep rheological data for a polydimethyl siloxane (PDMS) gel along with a fit to Equation \ref{eqn:chasset-thirion} is shown on a log-log plot in \textbf{Figure \textit{\ref{fig:soft_matter_adhesion}d}}, adapted from Reference \cite{karpitschka2015droplets}.
As crosslinking density increases, $\mu_0$ also increases, while $\tau_v$ decreases with an approximate scaling dependence $\tau_v \sim 1/\mu_0$ \cite{jeon2023moving,bantawa2023hidden,cass2024thesis}.
Interestingly, even significant changes in $\mu_0$ and $\tau_v$ have little effect on the power law exponent $n$, as this depends on the viscous rather than elastic properties of the material.

\begin{marginnote}[]
\entry{$G$}{Shear modulus, also commonly represented as $\mu$; for isotropic elastic materials, $G = \frac{E}{2(1+\nu)}$}
\entry{$G^*(\omega)$}{Complex modulus measured in shear rheology with an applied oscillatory shear at frequency $\omega$}
\entry{$G'(\omega)$}{Storage (elastic) modulus, the real part of $G^*(\omega)$, representing the in-phase stress response, at zero frequency equivalent to the shear modulus, $G'(0)=G$}
\entry{$G''(\omega)$}{Loss (viscous) modulus, the imaginary part of $G^*(\omega)$, representing the out-of-phase  stress response due to viscous dissipation in the solid material}
\end{marginnote}

Because viscoelasticity arises from local rearrangements of the elastic network, the relaxation time $\tau_v$ is independent of the length scale on which a measurement is made \cite{hu2012viscoelasticity}.
The viscoelastic properties of soft adhesives have long been understood to play an essential role in energy dissipation, especially during peeling or separation of an adhesive contact \cite{shanahan1995viscoelastic,Long1996,creton2003pressure,persson2005crack,muser2022crack}.
The relaxation time $\tau_v$ sets a time scale over which a solid adhesive material can move out of contact,
and hence strongly affects the maximum forces and deformations during a separation process---especially important in practical applications such as removing temporary adhesives without damaging a substrate or breaking apart the adhesive layer, or minimizing pain during removal of an adhesive bandage \cite{muser2022crack}.

\subsubsection{Poroelasticity}

Poroelastic relaxation provides an additional dissipation and relaxation mechanism in gels, and results from migration of the viscous fluid phase through the porous structure of a gel’s solid elastic network in response to elastic stresses \cite{hu2010using,hu2012viscoelasticity}. 
Unlike viscoelasticity, poroelasticity depends on the measurement length scale, because the poroelastic time scale $\tau_p$ is set by long-range migration of the fluid.
The effective diffusivity, $D^*$, of the fluid phase through the elastic network is given by  
\begin{equation}
    D^* = \frac{2(1-\nu)}{1-2\nu} \frac{G k}{\eta}
\end{equation}
\noindent where $\nu$ is the Poisson ratio, $G$ is the shear modulus, $k$ is the permeability of the network (with dimensions of area), and $\eta$ is the dynamic viscosity of the fluid \cite{hu2010using,xu2020viscoelastic}.
Note that although many gels and elastomers used in soft adhesion experiments have Poisson ratios close to 1/2, a finite compressibility of the elastic network ($\nu < 1/2$) is essential for poroelastic flow.
If a poroelastic material is deformed over a characteristic length scale $L$, the diffusive nature of poroelastic flow determines the time scale for poroelastic relaxation to be $\tau_p = L^2/D^*$, which grows quadratically with $L$.
Thus, poroelastic and viscoelastic effects can be distinguished by contact mechanics experiments with varied contact size \cite{hui2006contact,hu2010using,hu2011poroelastic,hu2012viscoelasticity}. 
Techniques have been developed for measuring the effective poroelastic diffusivity in soft gels using indentation methods, and were a major the topic of a recent comprehensive review of indentation of soft materials \cite{he2024comprehensive}. 

Multiple recent papers have found a key role for poroelastic relaxation in governing dynamic contact with soft gels, including investigations into the singular dynamics of adhesive detachment \cite{berman2019singular}, development of wetting-induced phase separation at the contact line \cite{qian2024emergence}, solvent migration near creases or folds in a soft gel \cite{flapper2023reversal}, adhesion with hydrogels \cite{lai2021relation}, and contact driven by poroelastic swelling---even to the point of fracturing the gel \cite{plummer2024obstructed,joshi2023energy}.
In other studies of dissipation mechanisms in soft gels, no single physical process was clearly dominant \cite{lai2019indentation,jeon2023moving}.
Significant recent efforts have sought to further develop theoretical frameworks for describing poroelastic mechanics in soft gels \cite{hu2012viscoelasticity,zhao2018growth,kopecz2023mechanical,he2024comprehensive}.

\section{PRACTICAL MATTERS: MEASURING SOFT ADHESION}
\label{sec:test_methods}

\begin{figure}[ht]
    \centering
    \includegraphics[width=\linewidth]{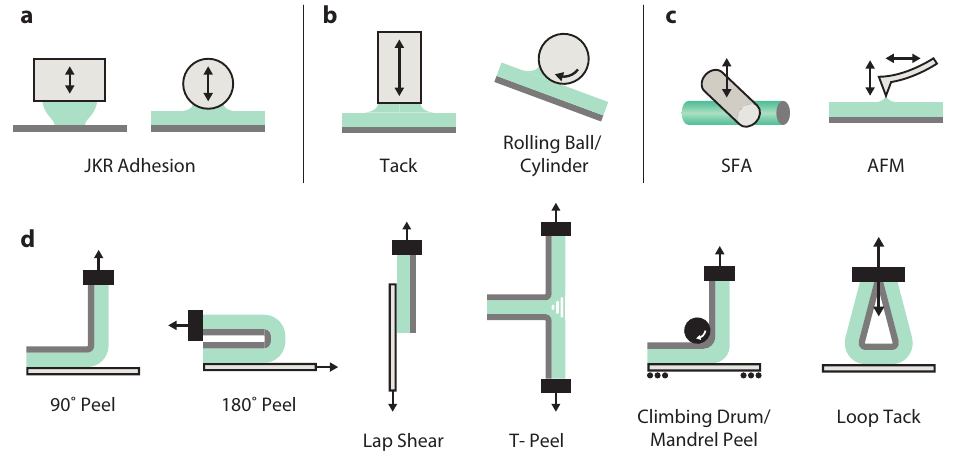}
    \caption{Testing soft adhesion. 
    \textit{(a)} JKR adhesion testing based on Equations \ref{eqn:JKR-a} and \ref{eqn:JKR-d}. 
    \textit{(b)} Standard probe-tack test and rolling ball comparative tack measurement. 
    \textit{(c)} Surface energy measurements using a Surface Forces Apparatus (SFA) or atomic force microscopy (AFM). 
    \textit{(d)} Tape peel adhesion tests. 
    Different geometries address different intended use cases for the adhesive.  
    Illustration created by Roderick W. Jensen, adapted with permission from a presentation created by Arnaud Chiche.
    }
    \label{fig:testing}
\end{figure}

We have seen that a complex interplay of surface and bulk properties determines the governing physics of how soft materials establish and maintain adhesive contacts over different length and time scales.
Within this complex parameter space are numerous opportunities to engineer adhesion at different scales and for different applications, as well as to use contact mechanics as a tool to measure and manipulate the fundamental properties of soft matter.
Here, we review some common methods for testing adhesion, as well as what mechanical and/or material properties can be measured using each test. 
Some of these tests are extremely simple and can be performed with minimal equipment, while others require more precise force, displacement, and imaging tools.

From a practical standpoint, measuring adhesion must consider how the adhesive will be used, examining not only the mechanical and material properties of the adhesive (such as modulus, viscoelasticity, and surface energy) but also the manner in which the adhesive contact will be formed and the interface loaded.
The compliance of the measurement system, the temperature and humidity of the laboratory, and the preparation of the countersurface can all greatly affect the measured adhesive properties.
The most useful real-world measurements usually come from treating the adhesion measurement not only as a quantification of the interface but also as an assessment of the adhesive system as a whole.

\subsection{Wetting and Contact Angle Tests}

A straightforward technique for characterizing the interactions of a liquid with a solid surface is a contact angle test, in which the angle formed between a small liquid droplet contacting a flat surface is measured visually in the configuration shown in \textbf{Figure \textit{\ref{fig:adhesion_wetting}b}}.
Practically, the measured contact angle is highly sensitive to surface inhomogeneities as the contact line moves and equilibrates, making it difficult or impossible to measure the thermodynamic Young-Dupr\'{e} contact angle $\theta$ discussed in Section \ref{sec:wetting} above.
Instead, dynamic contact angle measurements are used to measure the hysteresis or difference between an advancing contact angle $\theta_a$ as the liquid drop increases in volume and the receding contact angle $\theta_r$ as its volume decreases, or by examining the difference between the leading and lagging contact angles of a sliding droplet \cite{Gao2009Wetting, butt2022contact}.

\subsection{JKR Adhesion Testing}
\label{sec:JKR_test}

The most common contact mechanics method for fundamental adhesion measurements of soft solids is the ``JKR adhesion test'' or simply ``JKR test'' \cite{JKR1971,JohnsonBook1987,shull2002contact}, shown in \textbf{Figure\textit{ \ref{fig:testing}a}}. 
In this method, directly based on JKR contact mechanics theory (Equation \ref{eqn:JKR-a}), a spherical (or hemispherical) indenter is displaced toward a flat surface, contact is formed, and then the displacement of the probe is reversed until interfacial separation occurs.
Either the probe or the substrate can comprise the soft adhesive material, with the countersurface being rigid.
During a standard test, the displacement $d$ of the probe and the normal force $P$ at the interface are recorded while simultaneously imaging the circular contact area to obtain the contact radius $a$.
A JKR test can be performed under a constant displacement rate (more common) or constant loading rate and may or may not include a dwell time at the maximum compressive load before the probe is retracted.

Because the overall work of adhesion in such a test includes both thermodynamic and rate-dependent mechanical work, the results of a JKR test are described by Equations \ref{eqn:JKR-a}-\ref{eqn:JKR_pulloff_force} with a substitution of the strain energy release rate $\mathcal{G}$ for the thermodynamic adhesion energy, $W \rightarrow \mathcal{G}$ \cite{shull2002contact}.
Fitting with this version of these equations enables measurement of both the effective modulus $K$ of the soft material as well as $\mathcal{G} = \mathcal{G}(a)$ during the test.
In general, $\mathcal{G} \leq W$ during indentation and $\mathcal{G} \geq W$ during retraction, and thus the measurement provides bounds for the thermodynamic variable. 
The value of $\mathcal{G}$ near the start of retraction is identified as the critical strain energy release rate $\mathcal{G}_c$ that first allows the crack to open, and is usually what is reported from JKR tests as an upper bound estimate of $W$.

Earlier review articles describe this technique and associated data analysis in detail, including how to account for confinement effects in the case of thin samples \cite{shull2002contact, Hui2000Accuracy} as well as extensions that account for linear viscoelasticity \cite{shull2002contact,Lin1999Viscoelastic,Hui1998Contact}.

\subsection{Tack Tests}

Historically called ``quick stick'' \cite{dahlquist1966tack}, ``tack'' quantitatively measures how quickly an adhesive bond forms when two surfaces are brought together.
A ``probe tack'' adhesion test usually consists of a rigid cylindrical probe (or ``punch'') with a circular flat end that is brought into contact at a controlled rate with a flat, laterally extensive soft substrate as shown in \textbf{Figure\textit{ \ref{fig:testing}b}}. 
After contact is made, the surfaces are held in contact for some dwell time, and then separated.
The energy required to separate the surfaces is a quantitative measure of the tack for these contact conditions \cite{creton1996does,shull1998axisymmetric,hui2000mechanics}.
Due to high hydrostatic pressures that can develop within the adhesive layer underneath the probe, probe tack tests often reveal additional physics of the soft sticky material, especially for highly viscoelastic adhesives that can undergo instabilities and deformation mechanisms including cavitation, fingering instabilities, and fibrillation \cite{lakrout1999direct,josse2004measuring,crosby2000deformation,carelli2007effect}.

Other methods for determining relative tackiness have been developed in a wide range of industries.
For example, the rolling ball tack test, also shown in \textbf{Figure\textit{ \ref{fig:testing}b}}, is a common quality control measurement used to monitor drying rates of polymer coatings or compare tapes produced under different conditions \cite{Tse1999Rolling}.
In this technique, a ball is rolled down a ramp onto a flat substrate coated in the adhesive.
The further it rolls, the lower the tack.

\subsection{Surface Energy Tests}
\label{sec:surface_test}

Many other adhesion measurement systems have been designed to determine intrinsic interfacial properties like $W$ or $\gamma_{12}$. 
We show two examples in \textbf{Figure\textit{ \ref{fig:testing}c}}.

The Surface Force Apparatus (SFA) directly measures interfacial energy between two atomically smooth substrates, typically made of cleaved mica with surface coatings of interest \cite{tabor1969direct,israelachvili1972measurement,israelachvili1972NatMeasurement}. 
The substrates are curved into hemicylinders and brought into contact with their main axes at a right angle to one another, forming a circular contact area. 
Using interferometry and sensitive cantilevers, the SFA precisely measures displacements, interaction forces, and the area of contact, which can then be related to the interface properties.

A second highly sensitive method to measure contact mechanics at a small scale is atomic force microscopy (AFM).
While AFM is typically used to study the arrangement of atoms at a material surface, the separation force of the tip of an AFM probe from a contacting surface may also be used to determine nanoscale surface energy differences on a heterogeneous surface \cite{eastman1996adhesion, ferreira2010adhesion}. 
Further, micro- and nanoparticles can be mounted to AFM tips to enable direct measurement of the adhesion of interfaces too small to quantify with other standard methods \cite{harrison2015capillary}.

\subsection{Peel Tests}
\label{sec:peel}

Pressure-sensitive adhesives (PSAs) are the most common form of everyday soft adhesives, including tapes, reusable sticky notes, labels, and stickers. 
The adhesion of such structures is quantified with peel tests \cite{kaelble1969peel,yarusso1999quantifying}, which can be employed in a wide range of angles and geometries, as shown in \textbf{Figure\textit{ \ref{fig:testing}d}}.

In a typical angled peel test (e.g. 90$^\circ$ peel, 180$^\circ$ peel, or peel at an intermediate peel angle $\theta_p$), an adhesive coated on a backing layer is affixed to a rigid substrate. 
The backing layer is held at a fixed angle relative to the substrate, and the force $P$ required to peel the tape from the substrate is recorded.
The force is related to the energy per area  $\mathcal{G}$, the width of the tape backing layer $b$, and the peel angle $\theta_p$ by $P = \mathcal{G} b (1-\cos{\theta_p})$ \cite{Kendall1971Adhesion, Kendall1973Peel,Gent1977Peel}.
As in the JKR test, $\mathcal{G} \geq W$, as it reflects a combination of the thermodynamic adhesion energy $W$, the plastic bending energy, and the stretching energy of the backing layer.

Other common tests for PSAs shown in \textbf{Figure \textit{\ref{fig:testing}d}} include: lap shear, which measures force capacity of soft adhesives loaded in shear (like for reusable adhesive mounting hooks)\cite{bartlett2012looking}; 
T-peel, which measures deformation and adhesion of the adhesive with itself \cite{Hadavinia2006PeelPlast};
climbing drum and mandrel peel tests, which maintain a fixed bending radius of the tape \cite{kinloch1994peeling, williams2011mandrel, Kawashita2006Protocols});
and loop tack, which are often used for quality control measurements.

\section{SUMMARY AND OUTLOOK}

In this review, we have examined the fundamental physics of soft adhesion including thermodynamics, mechanics, materials science, and experimental methods.  
Many topics introduced only briefly could have been expanded into their own review articles; to the extent possible, we have pointed interested readers to additional references.
We also note that exciting connections exist between adhesion science and other closely related contact phenomena, such as friction and tribology, wetting and fluid droplet motion, and soft fracture mechanics.
While we focused our discussion on gels and elastomers as the most commonly used soft materials in adhesion experiments, the physics of soft adhesion can readily be applied across the vast field of soft matter.

New physics emerges as we explore contact with increasingly soft materials, 
from elastocapillary effects arising from surface tension competing with elasticity to complex interactions like poroelasticity and adhesion-induced phase separation.
Beyond developing our fundamental understanding of this extraordinary class of materials, 
experiments with soft matter often make length, time, and deformation scales accessible  that allow these materials to serve as model systems for exploring universal properties of traditional, stiff engineering materials.
We anticipate that research in the vital field of soft adhesion will continue to yield new insights and fundamental discoveries in the years to come.

\section*{DISCLOSURE STATEMENT}
The authors are not aware of any affiliations, memberships, funding, or financial holdings that
might be perceived as affecting the objectivity of this review. 

\section*{ACKNOWLEDGMENTS}
We thank Eric Dufresne, Robert Style, Frans Spaepen, Al Crosby, Costantino Creton, Anke Lindner, and many others---including the membership of the Adhesion Society---for countless helpful discussions over the years on the mechanics of interfaces and the science of soft adhesion. 
K.E.J. thanks Williams College for providing an extraordinary research environment, and the many past and current research students whose keen interest and endless questions directly informed and inspired the content of this review.
C.S.D. is grateful for the many mentors, collaborators, and students who have shaped her understanding of adhesion.
\textbf{Figure \textit{\ref{fig:testing}}} was adapted with permission from a course on ``Adhesion Characterization Methods'' created by Arnaud Chiche for DSM Research.
This material is based upon work supported by the National Science Foundation under Award No. CMMI-2129463.




\end{document}